\documentclass[sn-standardnature]{sn-jnl} 

\jyear{2025}%

\usepackage{aas_macros}
\usepackage{amsmath}
\usepackage{graphicx}
\usepackage{multirow}
\usepackage{bm}
\usepackage{siunitx}
\usepackage{comment}
\usepackage{fix-cm}
\usepackage{lmodern}
\usepackage{xr-hyper}
\usepackage{hyperref}
\usepackage{placeins}

\newcommand{\nablab}{\boldsymbol{\nabla}}
\newcommand{\Div}{\nablab\cdot}

\newcommand{\mul}{\eta_l}
\newcommand{\mus}{\eta_s}

\newcommand{\rhol}{\rho_l}
\newcommand{\rhos}{\rho_s}

\newcommand{\vl}{\mathbf{v}_l}
\newcommand{\vs}{\mathbf{v}_s}
\newcommand{\vbar}{\overline{\mathbf{v}}}
\newcommand{\Deltav}{\mathbf{\Delta}\mathbf{v}}
\newcommand{\taub}{\underline{\boldsymbol{\tau}}}
\newcommand{\taul}{\taub_l}
\newcommand{\taus}{\taub_s}
\newcommand{\taubar}{\bar{\taub}}

\begin{document}

\title[Lava Planets Interior Dynamics]{The role of interior dynamics and differentiation on the surface and atmosphere of lava planets}

\author*[1,2]{\fnm{Charles-\'Edouard} \sur{Boukaré}}\email{boukare@yorku.ca}

\author[3]{\fnm{Daphné} \sur{Lemasquerier}}\email{d.lemasquerier@st-andrews.ac.uk}

\author[4,5]{\fnm{Nicolas B.} \sur{Cowan}}\email{nicolas.cowan@mcgill.ca}

\author[2]{\fnm{Henri} \sur{Samuel}}\email{samuel@ipgp.fr}

\author[2]{\fnm{James} \sur{Badro}}\email{badro@ipgp.fr}

\author[6]{\fnm{Lisa} \sur{Dang}}\email{kha.han.lisa.dang@umontreal.ca}

\author[2,7,8]{\fnm{Aurélien} \sur{Falco}}\email{aurelien.falco@lmd.ipsl.fr}

\author[2]{\fnm{Sébastien} \sur{Charnoz}}\email{charnoz@ipgp.fr}

\affil*[1]{\orgname{Department of Physics and Astronomy, York University}, \orgaddress{\street{4700 Keele Street}, \postcode{M3J 1P3}, \city{Toronto}, \country{Canada}}}

\affil*[2]{\orgname{Université Paris Cité, Institut de physique du globe de Paris, CNRS}, \orgaddress{\street{1, rue Jussieu}, \city{Paris}, \postcode{75005}, \country{France}}}

\affil*[3]{\orgname{University of St. Andrews, School of Mathematics and Statistics}, \orgaddress{\street{North Haugh}, \city{St Andrews}, \postcode{KY16 9SS}, \country{UK}}}

\affil*[4]{\orgname{Department of Earth and Planetary Sciences, McGill University}, \orgaddress{\street{3450 rue University}, \city{Montréal}, \postcode{H3A0E8}, \country{Canada}}}

\affil*[5]{\orgname{Department of Physics, McGill University}, \orgaddress{\street{3450 rue University}, \city{Montréal}, \postcode{H3A0E8}, \country{Canada}}}

\affil*[6]{\orgname{Institut Trottier de recherche sur les exoplanètes, Université de Montréal}, \orgaddress{\street{1375 Av. Thérèse-Lavoie-Roux}, \city{Montréal}, \postcode{H2V0B3}, \country{Canada}}}

\affil*[7]{\orgname{Laboratoire AIM, Université Paris Cité, Université Paris-Saclay, CEA, CNRS}, \postcode{F-91191}, \city{Gif-sur-Yvette}, \country{France}}

\affil*[8]{\orgname{Laboratoire de Météorologie Dynamique, IPSL, CNRS, Sorbonne Université, Ecole Normale Supérieure, Université PSL, Ecole Polytechnique, Institut Polytechnique de Paris}, \orgaddress{\postcode{75005}, \city{Paris}, \country{France}}}

\newpage



\abstract{Lava planets are rocky exoplanets that orbit so close to their host star that their day-side is hot enough to melt silicate rock. Their short orbital periods ensure that lava planets are tidally locked into synchronous rotation, with permanent day and night hemispheres. Such asymmetric magma oceans have no analogues in the Solar System and their internal dynamics and evolution are still poorly understood. Here we report the results of numerical simulations showing that solid-liquid fractionation has a major impact on the composition and evolution of lava planets. We explored two different interior thermal states. If the interior is fully molten, the atmosphere will reflect the planet's bulk silicate composition, and the night-side solid surface is gravitationally unstable and constantly replenished. If the interior is mostly solid with only a shallow magma ocean on the dayside, the outgassed atmosphere will lack in Na, K and FeO, and the nightside will have an entirely solid mantle with a cold surface. We show that these two end-member cases can be distinguished with observations from JWST, offering an avenue to probe the thermal and chemical evolution of exoplanet interiors.}

\keywords{Exoplanets, Magma Oceans, Rocky Planets Formation and Evolution, Atmosphere-interior interaction.}

\maketitle


\newpage


\section*{Summary}

Lava planets are tidally locked exoplanets with molten daysides. Simulations show that their interior state—fully molten or with a shallow magma ocean—affects atmospheric composition and surface dynamics, and can be probed with JWST. 

\section{Introduction}

Astronomers have recently discovered a new class of exoplanets that offer a unique window into rocky interiors: lava planets \cite{leger2009,espinoza2021first}. 
These planets have densities consistent with an Earth-like bulk composition but orbits so close to their star that day-side equilibrium temperatures exceed the liquidus of silicates. Due to the tremendous tidal forces at these distances, lava planets should be tidally locked into synchronous rotation \cite{leger2009}. Tidal dissipation is an important heat source in lava planets, capable of maintaining their interior partially molten \cite{driscoll2015}. Unlike the relatively low-density short-period exoplanet 55 Cnc e,
\textit{bona fide} lava planets are expected to have lost all their volatiles to space \cite{valencia2010,owen2019atmospheric}, but their 2000--3000~K day-sides support an atmosphere of vaporized silicate rocks \cite{schaefer2007,castan2011,kite2016}, which may be observable with the James Webb Space Telescope (JWST) \citep{ito2015theoretical,nguyen2022impact, zilinskas2022}. Several studies used melt-vapor equilibrium model to predict the composition, dynamics, and spectral characteristics of lava planet atmospheres \cite{schaefer2009,zilinskas2022}. These approaches generally assume that the magma has the composition of the bulk silicate portion of the planet. However, these studies all noted that the magma ocean composition is in fact expected to differ from that of the bulk silicate portion of the planet because the magma ocean composition can evolve with time due to both liquid--vapor fractionation at the surface of the planet, and solid--liquid fractionation in the planetary interior \citep{kite2016}. In this study, we investigate the issue of the internal solid-liquid fractionation.

\section{Background}

Hemispherical magma oceans on lava planets challenge standard models of terrestrial interior dynamics and magma oceans \cite{kite2016}. A recent study explored the last stage of solid-state dynamics in lava planet interiors, in the presence of a global volatile atmosphere \citep{meier2023}. Here, we consider an alternative scenario with optically thin atmospheres on the day-side that are completely absent on the night-side \cite{castan2011,kite2016,nguyen2020}, consistent with the loss of a volatile atmosphere \cite{lammer2014origin,owen2019atmospheric}. Scaling laws have been used to predict first order aspects of lava planet dynamics with a day-side magma ocean \cite{kite2016}. Recently, significant advances have been made in understanding the fluid dynamics of the molten part of lava planet: the day-side magma ocean\cite{lai2024ocean1,lai2024ocean2}. Since supersonic winds are expected to blow from the day-side to the night-side over the magma ocean, these studies explored fluid dynamic regimes controlled by either wind forcing — where winds induce stresses at the top of the magma ocean —  or thermal forcing - thermal buoyancy decreases at the surface as the distance from the substellar point increases - , in addition to the effect of planetary rotation \cite{lai2024ocean1,lai2024ocean2}. However, these studies neglected the potential interaction between the magma ocean and the solid mantle, and focused on the late stages of lava planet evolution when the magma ocean is shallow and restricted to the day-side hemisphere. In this study, we aim to address this gap by investigating the silicate interior dynamics of lava planets as a whole, incorporating both the molten shallow MO, the solid mantle, and their chemical interactions. We will particularly focus on the long-term evolution of the interior, the internal chemical differentiation, and its link to atmospheric evolution, as this remains poorly understood in the context of lava planets\citep{castan2011,miguel2011,kite2016}.

Rock-forming elements such as Mg, Si, Na, or Fe have different affinities for solid \textit{vs.} silicate melts, leading to chemical fractionation. Incompatible elements prefer to remain in the melt, therefore their concentration in a magma ocean increases with ensuing solidification.
Solid-liquid segregation must be faster than magma ocean solidification such that silicate fractionation can occur \citep{chao2021}. Geochemical evidences of chemical fractionation in magma oceans have been found in our solar system for magma oceans \citep{rizo2013,blichert1999,bouvier2018} that crystallize in less than an hundred of millions years \citep{chao2021}. On a lava planet with a permanent day-side magma ocean, we expect long-term chemical fractionation of the magma ocean because gaseous, liquid and solid silicates are in contact for billions of years.  

\section{Numerical Simulations}


In this study, we are interested in the long-term evolution of lava planets over billions of years. We numerically modeled the interior dynamics of lava planets from the primordial fully molten state until its full solidification. We used two complementary codes to investigate (1) the solid-liquid multiphase fluid dynamics in 2D Cartesian geometry, (2) and the roles of spherical geometry, turbulence and rotation. Modeling parameters and methods are described in the section Methods of the main text (see also SI section 2 and 3). In the 2D simulations, we used the surface temperature gradient for the specific lava planet K2-141b accounting for the finite angular size of the star as seen from the planet, which means that more than half the planet is illuminated \cite{nguyen2020}. For the 3D spherical simulations, we used an idealized top temperature pattern where exactly half on the planet is illuminated.  We adopted the radius of K2-141b (1.5 R$_{\oplus}$) and a Core--Mantle Boundary (CMB) is located halfway between the center of the planet and its surface, consistent with the planet's bulk density \citep{2018A&A...612A..95B}. 

Additional aspects such as bulk composition, redox state, and fractional atmospheric escape are expected to play an important role in the chemical evolution of the magma ocean \cite{schaefer2009,curry2024evolution}. In this study, we focused on an unexplored first-order process which is a global differentiation in the planetary interior, induced by solid-liquid fractionation in the magma ocean. We coupled fluid dynamics and petrology in a self-consistent approach in order to highlight this process and its inner workings. To keep our message discernible and to focus solely on the consequences of internal dynamics, we chose to keep the chemistry as simple as possible. We consider the mantle as a closed system with the same composition of the bulk silicate Earth (BSE), i.e., SiO$_2$ 45.40 \%, MgO 36.76 \%
FeO 8.10 \% Al$_2$O$_3$ 4.48 \% CaO 3.65 \% Na$_2$O, 0.35 \% TiO$_2$ 0.21, and \% K$_2$O 0.03 \% (all in weight \%, see Supplementary Table 3), consistent with the composition of extrasolar rocky materials inferred from polluted white dwarfs \citep{Jura_Young_2014}. However, the  multiphase fluid dynamics simulation will track only the Fe/Mg in both solids and liquids (see Methods).

We followed the internal dynamical evolution of a lava planet chronologically from an initially hot and molten state \cite{chao2021} to the steady state stage when the mantle is mostly solid, except for the shallow day-side magma ocean \citep{kite2016}. Similarily to canonical magma oceans in our solar system \cite{chao2021}, we identified three stages: the global magma ocean, the mushy magma ocean, and the solid-state mantle. 

\subsection{Global magma ocean stage}

\begin{figure}
\centering
\includegraphics[width=0.8\textwidth]{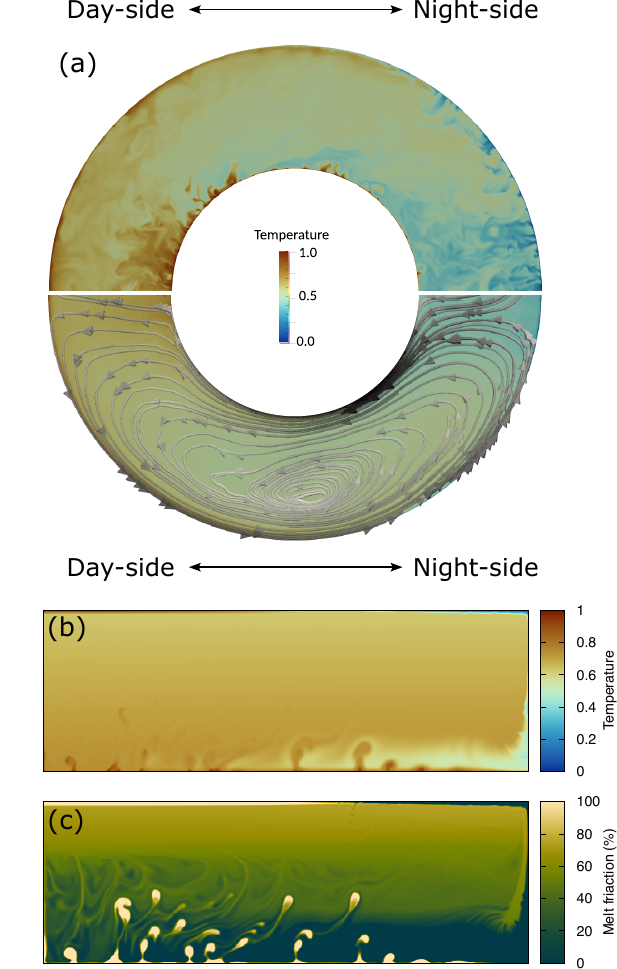}
\caption{\textbf{Numerical simulations of lava planets interior dynamics with no rotation.}  The dimensionless temperature scale is the maximum super-adiabatic temperature contrast: temperature is 1 at the base of the magma ocean, 0 at the surface on the night-side and 1 at the surface on the day-side. Panel (a) shows equatorial cross-sections from a turbulent model in spherical geometry with no phase change: the upper half is a snapshot and the lower half is a time average over 12 turbulent turnover times with superposed streamlines. Panel (b) shows temperature and panel (c) melt fraction from the 2D Cartesian model that accounts for solid--liquid phase change. See also Supplementary Video 1.  }\label{method}
\end{figure}

Our 3D simulations show that the interior temperature in the global magma ocean stage is homogenized by turbulent convective currents, although there remains a clear distinction between a colder and warmer hemisphere (Fig.~\ref{method}a). In our 2D simulations allowing for phase changes, however, the temperature follows that of the melting curve of silicates (because of latent heat release) \cite{fiquet2010}. In both 2D and 3D simulations, the imposed surface temperature gradient generates a global circulation, with downwellings on the night-side and a deeper return flow on the day-side (Fig.~\ref{method}a, see also Supplementary Figure 3 and 4). We refer to this persistent circulation, driven by the top boundary temperature gradients, as horizontal convection \cite{hughes2008}. In addition to horizontal convection, the internal dynamics of a lava planet is characterized by vertical convection driven by both the temperature difference between the surface and the interior, and the one between the interior and the metallic core. This vertical convection manifests itself as cold downwellings sinking in the night-side, and hot plumes forming at the CMB, in our 3D and 2D simulations (Fig.~\ref{method}a and \ref{method}b). Although the temperature difference between the sub-stellar point - that is, the location on the planet’s surface that is closest to the star - and the CMB might be locally below the thermal convection threshold, the low temperature of the night-side ensures that the magma ocean is always globally thermally unstable. 

\begin{figure}
\centering
\includegraphics[width=\textwidth]{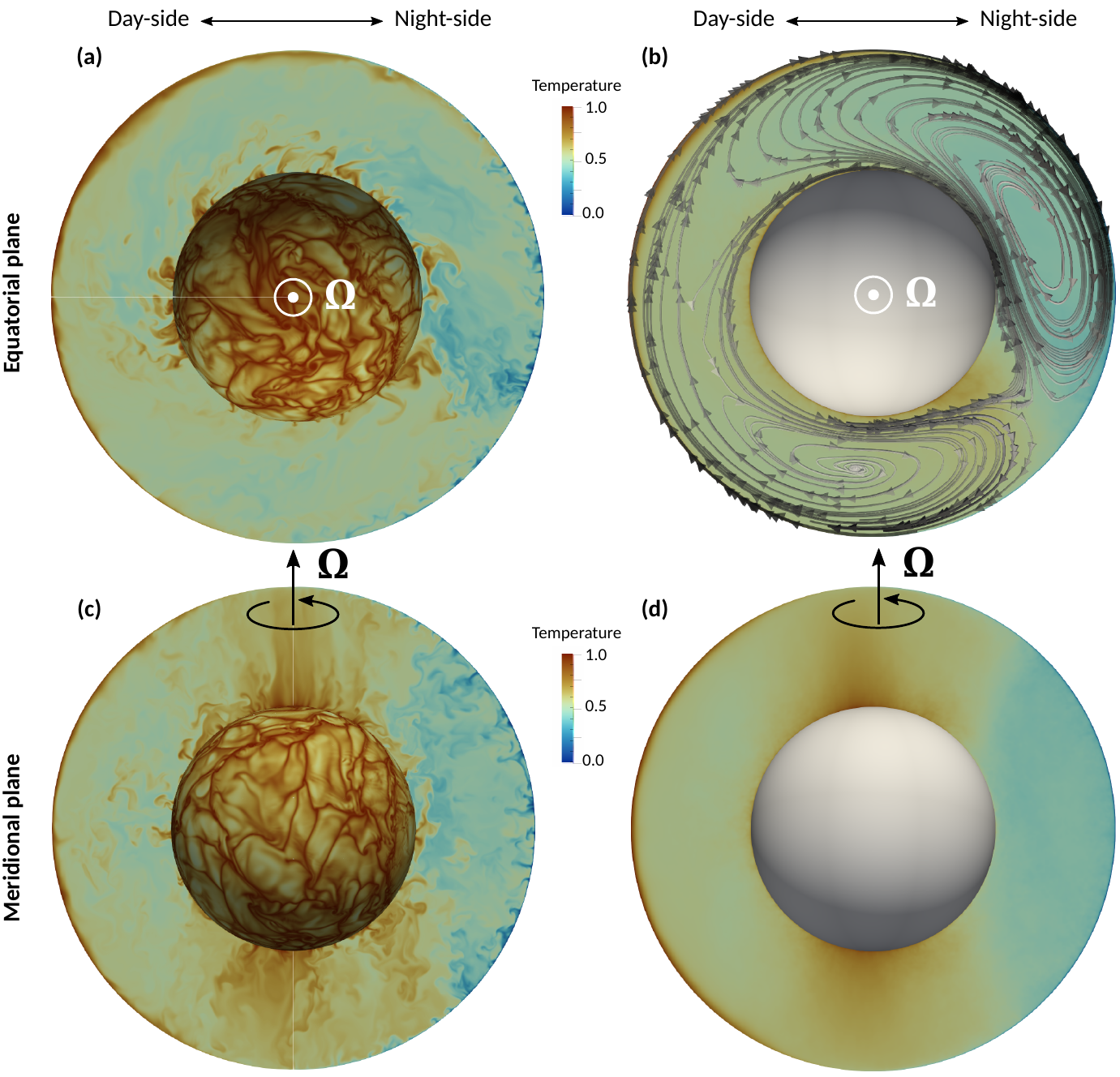}
\caption{\textbf{3D simulations of global magma ocean dynamics accounting for rotation.} Left panels show snapshots, while the right panels show time averages over 14 turbulent turnover times. The top panels show equatorial slices, while the bottom panels show meridional slices. In these simulations with a fully molten, turbulent magma ocean, rotation produces a longitudinal shift of downwelling and upwelling regions with respect to the anti-stellar and sub-stellar points, visible in panel (b). An asymmetry between the western and eastern hemispheres also develops in the bulk of the magma ocean.}\label{rotation}
\end{figure}

The global fully molten magma ocean stage is not expected to last longer than an hundreds years (see Supplementary Figure 6), consistent with canonical magma ocean evolution in our solar system \cite{lebrun2013}. For a low-viscosity global magma ocean, rotation could influence the dynamics \cite{maas2019} (Methods). The balance between buoyancy and Coriolis forces ---so-called thermal wind balance--- predicts radial flows to occur where the azimuthal temperature gradient is the largest \citep{zhang1992convection,dietrich2016core}. In regime where nonlinearity and viscous forces are neglected, this would lead to a 90° longitudinal shift of the mean circulation with respect to the sub-stellar and anti-stellar point. Non-linearities due to temperature advection and inertia, together with the development of strong zonal flows, complicate this simple picture by shifting the downwelling currents, and introducing an asymmetry between the warm and cold hemispheres (Fig.~\ref{rotation}a-b and Supplementary Figure 4). Note that the origin of this shift is \textit{a priori} different from the atmospheric hotspot offsets predicted by shallow atmosphere models where equatorially trapped waves play a prominent role \citep{showman2011equatorial}.

\subsection{Mushy magma ocean stage}

\begin{figure}
\centering
\includegraphics[width=\textwidth]{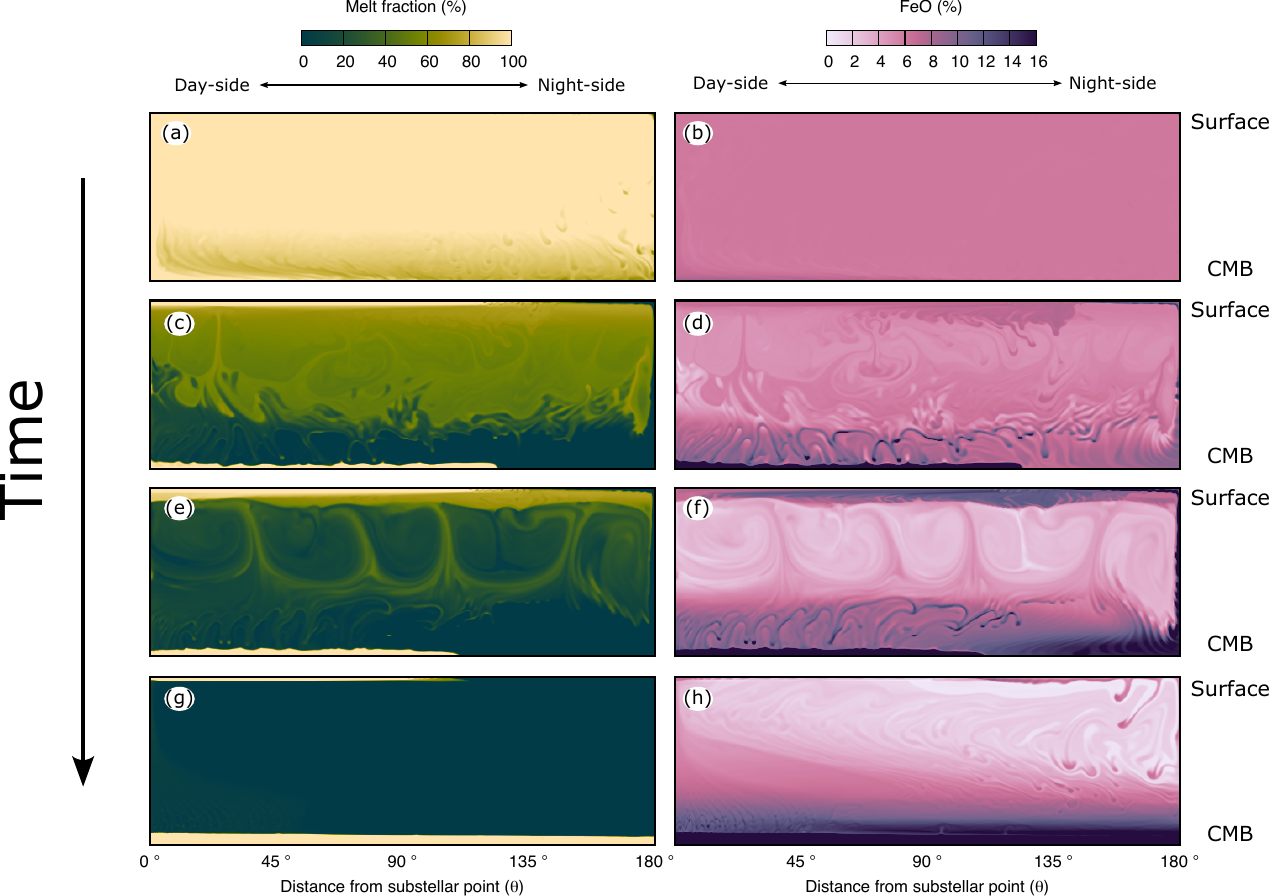}
\caption{\textbf{Thermal and chemical evolution of a lava planet's magma ocean.} Left panels shows the melt fraction, while right panels show the FeO concentration. The four snapshots in time starting from a molten, chemically homogeneous mantle (top panels), to a solid, and chemically differentiated, mantle (bottom panels).  Panels (a) and (b) show the initial conditions: the magma ocean is well mixed and nearly entirely molten. The composition of the magma ocean corresponds to the bulk silicate planet's composition,  \textit{i.e.}, 6 mol.\% of FeO (BSE composition). This corresponds to the hot end-member scenario depicted in Figure \ref{sketch}. Panels (c) and (d) show the onset of the mushy stage: the light green color in panel (c) represents a 50/50 mix of melt and solid, while the dark magenta in panel (d) shows iron-rich solids accumulating in the solid deep mantle,above the core-mantle boundary. Concomitantly, an iron-rich shallow magma ocean starts forming below the surface between the sub-stellar point and 120 degrees. Panels (e) and (f) show the end of the mushy stage: the FeO concentration of the shallow day-side ocean magma reaches its maximum (12 mol.\% FeO), and lies above a lighter and essentially solidified mantle. Panels(g) and(h) show the solid stage; the iron-rich shallow magma ocean has been buried in the deep mantle by gravitational overturn. The remnant day-side shallow magma ocean has a FeO concentration of only about 2\%. This corresponds to the cold end-member scenario depicted in Figure~\ref{sketch}.}
\label{time_evolution}
\end{figure}

Starting with an initial global magma ocean that is compositionally homogeneous (Fig.~\ref{time_evolution}a--b), both 2D multiphase and 3D fully molten simulations yield similar results: a statistically steady circulation is observed, with downwellings on the night-side, and a global return flow to the day-side (see Supplementary Figure 3). Our 3D spherical simulations confirm this even in the presence of turbulence and rotation, rendering the circulation pattern robust, and validating the 2D Cartesian approach (see Supplementary Figure 3).

Upon cooling, the solid fraction increases to the point where the silicate interior behaves more like a solid than a liquid (Fig.~\ref{time_evolution}c-d); this rheological transition is expected to occur around 40\% melt fraction \cite{Lejeune&Richet95}. This marks the appearance of a mushy region  where the melt fraction ranges between 40\% to 60\% \cite{Lejeune&Richet95}. The mushy layer grows rapidly to occupy the entire mantle, consistent with classical magma ocean thermal evolution models \cite{lebrun2013}. We define the onset of the mushy stage when the rheological front (40\% melt fraction) reaches the surface of the planet (Fig.~\ref{time_evolution}c--d). With no additional heat source such as tidal heating, the mushy stage may last up to hundreds of millions years (see Supplementary Figures 6) \cite{lebrun2013,herath2024thermal}. 

During this mushy stage, solids and liquids segregate, differentiating the planet (Fig.~\ref{time_evolution}e--f). FeO being an incompatible component \cite{nomura2011}, it prefers to remain in the melt rather than  in solids. The shallow day-side magma ocean FeO composition increases up to 12~mol.\%, twice the initial value (BSE) (Fig.~\ref{time_evolution}f).  The FeO concentration can increase in the shallow magma ocean, which remains gravitationally stable, as long as these FeO-rich silicates remain molten. However, horizontal convection transports these iron-rich melts towards the night-side, where they solidify. FeO-rich solids are substantially denser than the surrounding mantle and therefore sink to the deep mantle. Consequently, the FeO-rich shallow day-side magma ocean is progressively replaced by a FeO-poor magma ocean (Fig.~\ref{time_evolution}g). The role of phase segregation efficiency on producing a chemically stratified planet is discussed in SI Section 7 (see Supplementary Figure 7).    
\subsection{Solid-state mantle stage}

Lava planet enters its long lasting solid-state stage when the lava planet becomes essentially solid (Fig.~\ref{time_evolution}g and Fig.~\ref{time_evolution}h, see also Fig.~S6). The remaining day-side magma ocean is shallow, less than 200~km thick, because the cooling from the night-side has efficiently cooled the entire planet interior, not only its night-side hemisphere. This is at odds with the results in \cite{boukare2022} where  heat transport between the two hemispheres of the planet was neglected.  Even if the core is relatively hot (see Methods), a lava planet with no persistent heat sources will almost completely solidify. When the lava planet start its long-term solid-dynamics,  the entire mantle is compositionally stratified, affecting solid-state convection. Dense geochemical reservoirs may never be exposed to the surface. With a colder solidification temperature than Mg-rich silicate, the FeO-rich lowermost mantle remelts (Fig.~\ref{time_evolution}g-h), forming a basal magma ocean, as suggested for the Earth \citep{labrosse2007}, Mars \cite{samuel2021thermo} and Venus \cite{o2020venus}.
\\

\section{Observability}

\begin{figure}
\centering
\includegraphics[width=\textwidth]{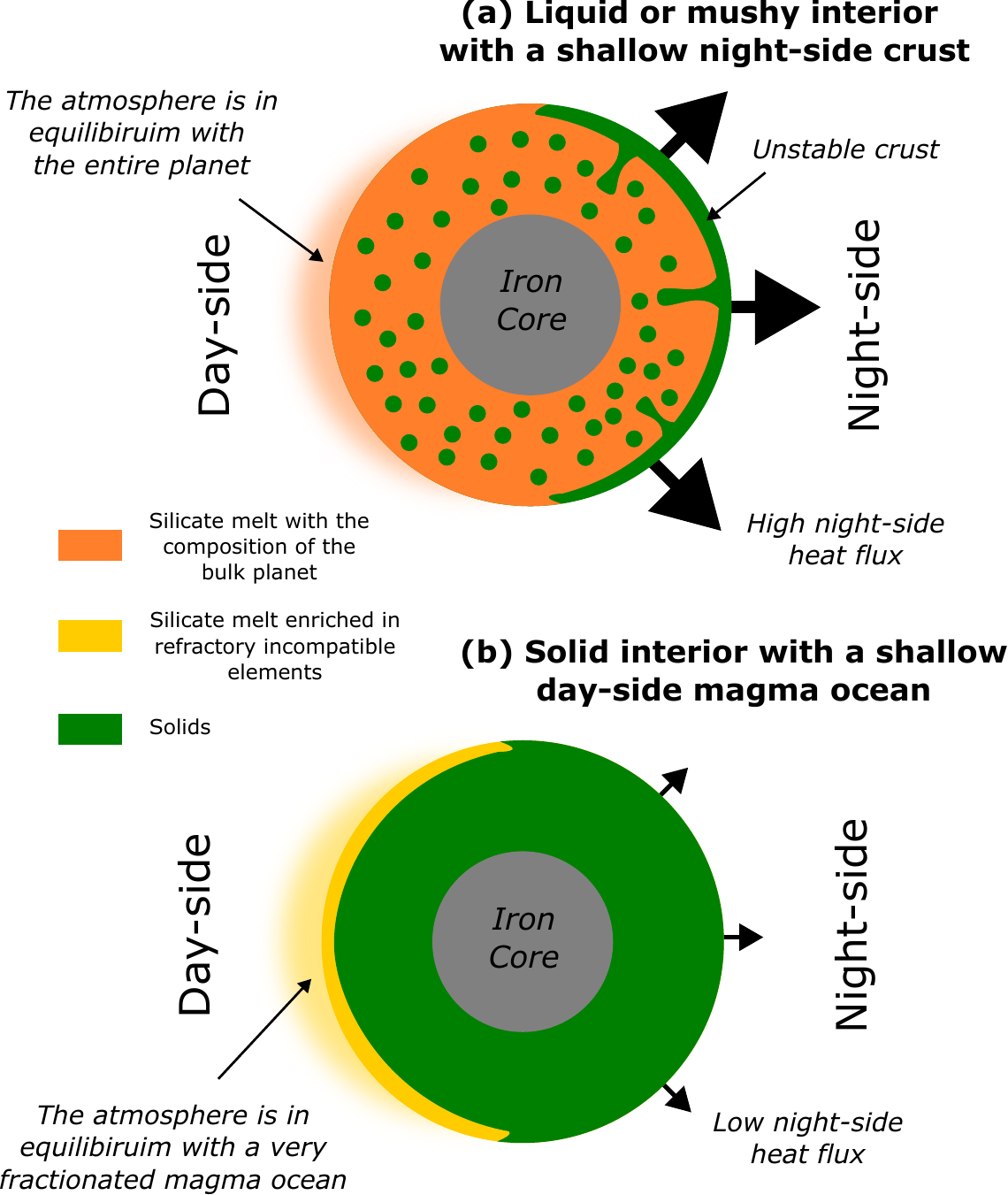}
\caption{\textbf{Stages of lava planet magma ocean for a hot \textit{vs.} cold interior.} Panel (a) shows the hot end-member, characterized by an essentially liquid interior (orange, see also Fig.\ \ref{time_evolution} panels a and b). Vigorous convection in the liquid state favors compositional mixing and efficient cooling by heat transfer from the day-side to the night-side hemisphere. Panel (b) shows the cold end-member, characterized by a solid interior (green, see also Fig.~\ref{time_evolution} g and h) with a shallow day-side magma ocean. Solid-liquid gravitational segregation has chemically differentiated the magma ocean (yellow). In this case, solid-state convection is too weak to generate an observable thermal signature on the night-side surface. }\label{sketch}
\end{figure}

Lava planets have been identified as target of interest to be observed with JWST: with five different programs to observe lava planets \citep{hu2021determining,brandeker2021raining,dang2021hell,espinoza2021first}. Given their short orbital periods, lava planets are particularly amenable to full-orbit phase-resolved spectroscopy. In the light of our results, we postulate two observable end-member stages for lava planets: either a hot, homogeneous global magma ocean stage, or a cold, chemically differentiated solid-state stage (Fig.~\ref{sketch}). The hot and cold stages can be distinguished by measuring the infrared continuous full-orbit phase curve and the day-side and night-side emission spectra of a lava planet. 

If the magma ocean interior is hot and essentially molten, the night-side is expected to radiate at a high temperature. The unstable gravitational solid surface is constantly replenished by hot lava that radiate at about 1500~K before solidifying and being replaced. If the magma ocean extent is global, the heat flux coming from the interior due to vertical convection is expected to be very high, \textit{i.e.}, larger than 10$^3$~W~m$^{-2}$, and it could be as large as 10$^6$~W~m$^{-2}$ \citep{lebrun2013}. Similar values are obtained using scaling laws relevant for horizontal convection \cite{hughes2008}. Endogenic heat fluxes of $10^3$~W~m$^{-2}$ and 10$^6$~W~m$^{-2}$ correspond to a night-side surface temperature of 900~K and 3000~K, respectively (see SI section 6 and Supplementary Figure 5). With no atmosphere on the night-side, a hot night-side surface temperature on a lava planet  is an univocal signature of a hot internal state of the planet. A thin rock vapour atmosphere above the magma ocean would not extend to the planet's night-side and could not transport significant heat \cite{castan2011}. Alternatively, a thick, volatile-rich global atmosphere could advect heat to the night-side \cite{hammond2017}, but such a scenario is not expected due to the extreme atmospheric loss \cite{Ito_Masahiro_2021} and in any case would betray itself via the planet's emission spectrum \cite{2023ApJ...954...29P}. Conversely, a very viscous solid interior cannot advect significant heat to the night-side, resulting in associated temperatures of the order of 100~K \cite{kite2016}.

\begin{figure}
    \centering
    \includegraphics[width=0.5\textwidth]{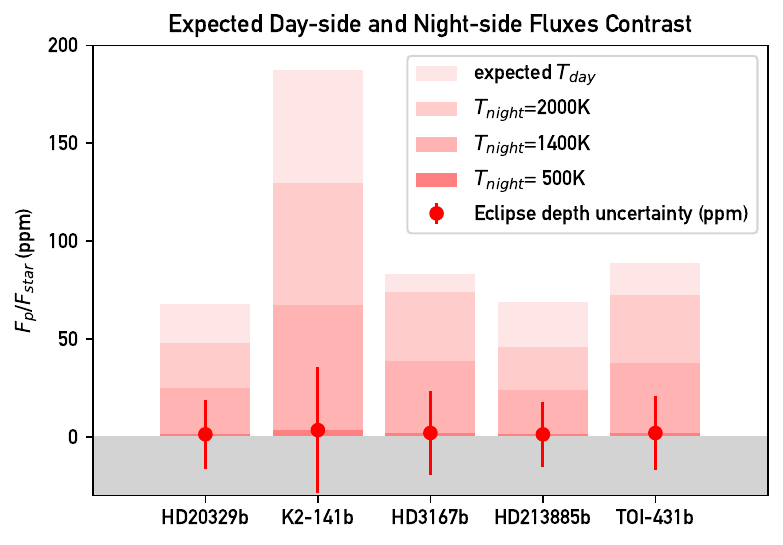} \\
    \includegraphics[width=0.5\textwidth]{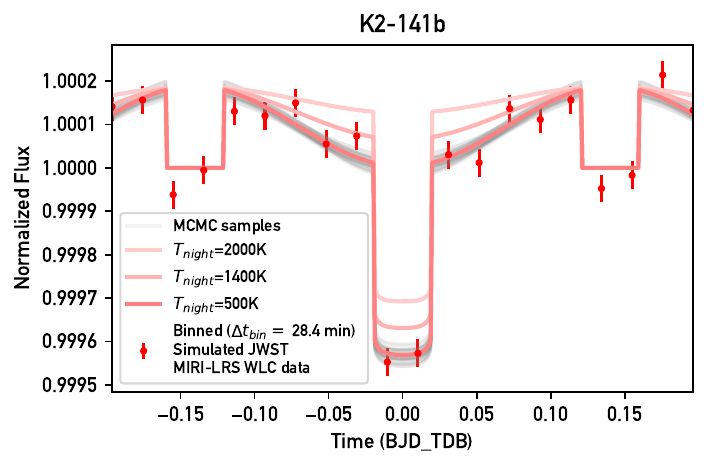} 
    \caption{\textbf{Simulated JWST constraints on the nightside temperature of lava planets.} The top panel shows the expected day-side flux and 3 different night-side fluxes to demonstrate the expected phase curve amplitudes for different lava planets. We note that the night-side flux for the 500K cases is only a few ppm for most cases. The error bars show the photometric uncertainty achieved for a temporal bin of the duration of the eclipse for each planet, respectively. The bottom panel shows different phase curves models of a known lava planet, K2-141b, with 3 different night-side temperatures to demonstrate that it is possible to distinguish a cool night-side from a hot night-side. Overlain is an MCMC fit to phase-folded simulated JWST/MIRI/LRS white-light phase curve measurements for a night-side temperature of 500~K. The binned data points and uncertainties assumes that 2 continuous full-orbits of the planet are observed and have evenly-spaced temporal bins of $\Delta t_{\rm bin}=$28.2 minutes.}
    \label{fig:phasecurvemiri}
\end{figure}

The JWST can probe the night-side temperature through mid-infrared full-orbit phase curve observations of transiting lava planets. Due to the large angular size of the star from the planet’s point of view, the planet can sometimes be illuminated beyond 90 deg from the substellar point. In the case of K2-141b, with no other heating source, the night-side illumination is predicted to be only on the order of 10~ppm and a disc-integrated average night-side temperature of 500~K \cite{nguyen2020}. We show in Figure~\ref{fig:phasecurvemiri} that we can distinguish an internally heated night-side from a cold night-side by their phase curve amplitudes (see also Supplementary Figure 10). We also calculate the expected flux difference between the expected day-side and different night-side temperatures for some of the most favourable lava planet targets along with their expected eclipse depth uncertainty (see Methods, section 4.4 for details). 

Moreover, in the hot-interior scenario, the magma ocean and its overlying atmosphere reflect the bulk silicate composition of the planet, so we expect emission spectroscopy to reveal Na, K and FeO. On the other hand, if the planet's interior has mostly solidified,  we expect relatively volatile Na and K to be locked in night-side glaciers \cite{nguyen2020} and FeO to be trapped in the deep mantle, leading to an atmosphere that will only exhibit SiO spectral features. In the cold scenario, the low condensation temperature of Na, K and FeO allows them to condensate beyond the shore line of the day-side magma ocean. They are then
cold-trapped in the permanently dark regions of the planet. In the warm interior scenario, the night-side surface is constantly floundering and remixing with the magma ocean, replenishing the supply of alkali elements to evaporate at the sub-stellar region. We calculated the atmospheric composition above the magma ocean by solving the chemical equilibrium of a gas evaporated by a melt with Bulk Silicate Earth (BSE, \cite{mcdonough1995}) composition (as in \cite{wolf2022vaporock,van2022lavatmos,charnoz2023effect}), assuming local temperature equal to planet's equilibrium. Two melt compositions were considered: a case with full BSE composition, and an evolved BSE case devoid of K, Na and Fe (see Supplementary Table 3).

\begin{figure}
    \centering
    \includegraphics[width=0.8\textwidth]{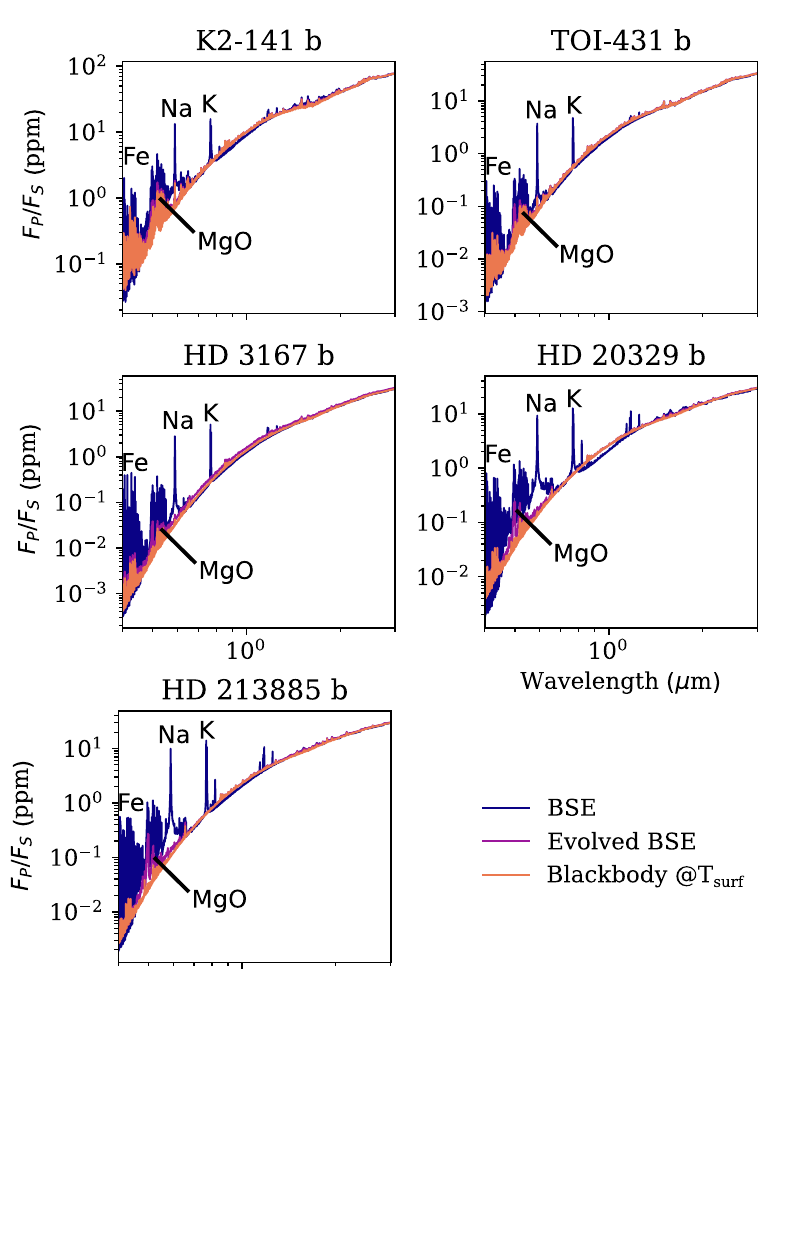}
    \caption{\textbf{Simulated emission spectra for five potential lava planet targets.} Each panel shows the simulated ratio of dayside planetary to stellar flux ---the eclipse depth--- in parts-per-million for three chemical scenarios. The planetary spectra were calculated with ATMO \citep{ tremblin2016cloudless} between 0.4 and 3.0~$\mu m$ (log-scale).
    Blue lines are for a BSE magma composition, magenta lines denote an evolved BSE (no Na, K and Fe), while orange lines are the null hypothesis of a blackbody planetary spectrum.
    The BSE and evolved BSE compositions are given in the Supplementary Table 3).  At these wavelengths, the strongest spectral features of the BSE scenario are Fe, Na, and K.  Those features are absent from the evolved BSE spectrum, however, leaving only MgO. The remaining small spectral features are due to the star, not the planet and hence also appear in the blackbody scenario.  
    }
    \label{fig:spectra}
\end{figure}

The resulting emission spectra at the day-side (secondary eclipse), shown in Fig.~\ref{fig:spectra} have been computed using ATMO \citep{tremblin2016cloudless}, an atmospheric code that solves for radiative transfer and local thermodynamical equilibrium in a hydrostatic atmosphere. In the BSE case, Na and K spectral features, at 0.6 and 0.8~$\mu m$, respectively appear prominently. Fe is preponderant for wavelengths~$< 0.6~\mu m$, too short to be observed with the James Webb Space Telescope (JWST), but potentially observable with the Habitable Worlds Observatory planned by NASA for the 2040s. This presence of emission features is due to the Pressure-Temperature (PT) profile of the atmospheres (see Supplementary Figure 8), which present a strong thermal inversion (the temperature increases with the altitude). The vertical temperature profile of lava planet atmospheres also depend on poorly-constrained atmospheric dynamics \citep{nguyen2022impact} and hence could be stronger or weaker than this 1D model. Using an evolved BSE composition, the Na and K features unsurprisingly disappear (the PT profiles are closer to isothermal and thus present less emission features). Depending on the strength of the day-side vertical temperature profile inversion, K and Na emission features could in principle be detected by JWST/NIRISS/SOSS \citep{2023Natur.614..670F}. MgO and SiO emission features (9 and 15 $\mu m$) are present but weak (see Supplementary Figure 9). The evolved BSE case shows no significant feature and could be compatible with a black-body emission surface in the displayed spectral range. Discriminating between the two cases would therefore be challenging using JWST's MIRI spectrometer (5--29 $\mu m$). The challenge of distinguishing between BSE and evolved magma ocean chemistry on a lava planet is comparable to that of detecting atmospheres on potentially habitable worlds orbiting red dwarf stars \cite{2019AJ....158...27L}. Lastly, we note that although transit spectroscopy of the hydrostatically bound atmosphere is onerous \citep{nguyen2022impact}, the escaping exosphere of a lava planet may be amenable to transit spectroscopy, as has been done for larger exoplanets \citep{2015Natur.522..459E}. 


For old planetary systems, mantle temperature can further be linked to tidal forces and dissipation in the planet. If a lava planet orbiting a 10 Gyr-old star is found to be in the hot interior state, then the interior has to be tidally heated. This would extend the lifetime of the partially molten interior, because tidal dissipation is maximum in partially molten contexts \cite{kervazo2022inferring}, and provides an energy source to continuously reheat the planet. This process would fail for an entirely molten planet, where tidal heating is small, resulting in rapid cooling and bringing the planetary interior to a partially molten state. Interesting feedbacks or runaway effects may emerge from the coupling of thermal interior evolution and tidal dissipation. 

\section{Conclusion}

We investigated the multiphase fluid dynamics of lava planets over billion-year timescales, from their formation to the point where they achieve a thermal (pseudo) steady-state. Lava planets are expected to be mostly molten immediately after their formation. Despite being heated from the top on the day-side, they solidify almost as quickly as magma oceans in our solar system. However, unlike rocky planets in the solar system, lava planets maintain a long-lived, shallow magma ocean on their day-side, even after billions of years of interior cooling (see also \cite{herath2024thermal}).

Rock-forming oxides such as FeO, Na$_2$O, and SiO$_2$ exhibit different affinities for the solid and liquid silicate phases. During the solidification of the global magma ocean, and at the shorelines of the long-lived shallow magma ocean, solids crystallize from the melt. Since solids with distinct compositions can segregate from the melt, our study reveals that lava planets undergo solid-liquid chemical fractionation throughout their entire existence. As a result, the silicate atmosphere of a relatively old lava planet should be in equilibrium with a chemically processed magma ocean whose composition differs from the bulk silicate portion of the planet.

Moreover, the night-side surface temperature of a young lava planet is expected to be relatively high, around 1500 K, to radiate the heat supplied by convective motions in the planet’s interior. However, without an additional heat source, the night-side surface temperature of an older lava planet is expected to be significantly cooler. The present-day thermal state of a planet is a direct consequence of its entire thermochemical history since formation. Therefore, probing the mantle temperature of these distant rocky worlds may reveal fundamental aspects of planetary evolution.

We demonstrated that measuring the night-side surface temperature of lava planets is currently within the capabilities of telescopes such as JWST, offering promising constraints on their internal thermal state. Future ground-based observations, such as those from the ELT, which should be able to characterize the composition of lava planet silicate atmospheres, may allow us to test intriguing thermo-chemical couplings between the atmosphere, silicate melt, and silicate minerals in the planet’s interior, as suggested in this study.


\section{Methods}

\subsection{2D multiphase flow model}

We model the solid-liquid multiphase flow dynamics using the code {\tt Bambari} \citep{boukare2017}.  The code implements the multiphase flow mathematical formalism based on the averaging method \cite{drew1971,mckenzie1984,bercoricard2001,boukare2017,keller2019} where two two mechanical phases are considered: a liquid phase associated with a melt fraction $\phi$ interacts with a deformable solid matrix whose complementary fraction is $1-\phi$. The original numerical implementation \cite{sramek2007,sramek2007these} of the set of governing equations was considerably improved in recent years, in particular with the use of the numerical stencils implemented for the momentum conservation in the Finite-Volume code {\tt StreamV} \cite{Samuel2012,Samuel2018}. The latter allow handling reliably large and sharp viscosity contrasts with negligible spurious pressure effects \cite{Samuel&Evonuk2010}. We have also extended the model of \cite{boukare2017} from two to four phases to capture chemical fractionation between liquid and solid during phase change. 
Each mechanical phase is now composed of two major compositional phases that correspond to a FeO-rich and member and a MgO-rich end member. In addition to the advection-diffusion of temperature, our model tracks the advection of 4 compositional fields : 2 liquids (the MgO-rich end-member and FeO-rich end-member) and the 2 solids counterparts.

Mass conservation for the four major chemical components (one MgO-rich end-members, and one FeO-rich end-members for the solid and the liquid mechanical phases) write:

\begin{equation}
\begin{split}
      \frac{\partial \phi_1}{\partial t} + \Div \vs \phi_1 = - \Gamma_1 \\
      \frac{\partial \phi_2}{\partial t} + \Div \vs \phi_2 = -\Gamma_2 \\
      \frac{\partial \phi_3}{\partial t} + \Div \vl \phi_3 = +\Gamma_1 \\
      \frac{\partial \phi_4}{\partial t} + \Div \vl \phi_4 = +\Gamma_2, \\
\end{split}
\end{equation}
where $\vl$ is the velocity vector of the liquid phase, $\vs$ is the velocity vector of the solid phase and $\Gamma_i$ are associated with the rate of phase change. $\phi_1$ and $\phi_3$ are the fraction of the MgO-rich end-member in the solid, and liquid phase, respectively. $\phi_2$ and $\phi_4$ are the fraction of the FeO-rich end-member in the solid, and liquid phase, respectively. The densities of the four chemical components are:

\begin{equation} 
	\begin{split}
	\rho_1 = \rho_0(1-\alpha T)+\frac{1}{2} \Delta \rho_1 -\frac{1}{2} \Delta \rho_2  \\
	\rho_2 = \rho_0(1-\alpha T)+\frac{1}{2} \Delta \rho_1 + \frac{1}{2} \Delta \rho_2 \\
	\rho_3 = \rho_0(1-\alpha T)-\frac{1}{2} \Delta \rho_1 -\frac{1}{2} \Delta \rho_2  \\
	\rho_4 = \rho_0(1-\alpha T)-\frac{1}{2} \Delta \rho_1 +\frac{1}{2} \Delta \rho_2,  \\
	\end{split} 
\end{equation} 
where $\rho_0$ is the reference density, the thermal expansivity $\alpha$ is considered constant in the solid and liquid phase and $T$ is the temperature. The chemical density contrast between the dense and light end-members is $\Delta \rho_2$. The isochemical density contrast between the liquid and solid, $\Delta \rho = \rhos-\rhol$  is equal to $\Delta \rho_1$. We parameterized these density contrasts using the thermodynamic model of  \cite{boukare2015} (see Supplementary Figure 1).

Assuming local thermal equilibrium between all phases, (\textit{i.e.}, the temperature in all the phases are the same), the thermal state is described by one energy conservation equation:
\begin{equation}
	\partial_t T + \vbar \cdot \nablab T  = \nablab^2 T - (\Gamma_1+\Gamma_2)S_t,
	\label{eq:energ_cons}
\end{equation}
where $\vbar$ is the velocity of the solid-liquid mixture, $S_t$ is the dimensionless Stefan number:
\begin{equation}
 S_t = \frac{L}{C_P \Delta T_m}, 
\end{equation}
where $C_P$ is the effective heat capacity of the mixture of the four components, $L$ is the latent heat release during solid-liquid phase change, and $\Delta T_m$ the reference temperature difference.

The velocity of the solid-liquid mixture, $\vbar$, is related to the solid, $\vs$, and liquid, $\vl$, velocities by the following equation: 
\begin{equation}
    \vbar = \phi \vl + (1-\phi)\vs .
\end{equation}

The conservation of momentum of the average solid-liquid mixture in dimensionless form writes:
\begin{equation}
    -\nablab \Pi + \Div{\taubar} - Ra T \bm{e}_g  - Rp \phi \bm{e}_g + \frac{1}{2} Rc (\phi_2+\phi_4-\phi_1-\phi_3) \bm{e}_g  = 0
\end{equation}
where $\nablab \Pi$ is the dynamic pressure gradient, $\taubar$ is the stress tensor of the solid-liquid mixture, and $\bm{e}_g $ is a unit vector pointing upwards. 
The thermal Rayleigh number, $Ra$, and respectively the phase and compositional buoyancy numbers, $R_p$ and $R_c$ are:
\begin{equation}
	\begin{split}
		Ra=\frac{\rho_0 \alpha g \Delta T_m g D^3}{\kappa \eta_s}, \\
		Rp=\frac{ \Delta \rho_1 g D^3}{\kappa \eta_s}, \\
		Rc=\frac{ \Delta \rho_2 g D^3}{\kappa \eta_s}.
	\end{split}
\end{equation}
where $\Delta \rho_1$ and $\Delta \rho_2$ are the isochemical density contrast and maximum chemical introduced above. The phase buoyancy number $R_p$ is associated to the density difference between melt and solid of same chemical composition. The compositional buoyancy number $R_c$ describes the effect of composition, \textit{i.e.}, FeO content, on density. Both buoyancy numbers are depth-dependent and parameterized based on a self-consistent thermodynamic model of \cite{boukare2015} (see SI section 1). We define the stress tensor of the solid-liquid mixture as:
\begin{equation}
    \taubar = (1-\phi) \taus + \phi \taul = \eta (\phi)\left(\nablab \vbar+ \left[ \nablab \vbar \right]^t \right),
\label{eq:taubar}
\end{equation}
where $\eta(\phi)$ is the viscosity of fluid mixture that varies from $\eta(\phi = 0) = \mus$ to $\eta(\phi = 1) = \mul$. The viscosity of the mixture, therefore only depends on $\phi$, and we neglect its dependence with pressure and temperature.
This simplification is motivated by the fact that the influence of melt-solid viscosity contrast dominates over pressure and temperature effects \cite{Lejeune&Richet95,karato1993rheology}.
Temperature- and pressure-dependent viscosity can be reasonably approximated by a homogeneous viscosity case with a representative effective viscosity (\textit{i.e.}, relying on a volume-averaged viscosity) \cite{solomatov1997three}. 
The viscosity ratio  between the solid and the melt phase is set to 1000, which is sufficient to decouple the rheology of the solids and liquids. Larger solid-melt viscosity contrasts are therefore not expected to affect significantly our results and conclusions.

To distinguish between the solid and the liquid phase velocities, we use a retro-action equation (see \cite{bercoricard2001,boukare2017}) that quantifies the velocity difference between the two phases, $\Deltav = \vs - \vl$: 

\begin{equation}
\begin{split}
      \phi \Deltav = \phi^2 \delta^2 (1-\phi)
	\left(\nablab \left[ \frac{(1-\phi)}{\phi} \Div \phi \Deltav \right] 
	+  X\bm{e_g} 
	 + \Div{\taubar}
	 \right), 
\end{split}
\label{eq:retroaction}
\end{equation}
with
\begin{equation}
    X = Rp + \frac{1}{2}\left(\frac{\phi_3-\phi_4}{\phi}+\frac{\phi_2-\phi_1}{1-\phi}\right)Rc,
\end{equation}
$\delta$ is the melt mobility number, and $\zeta$ is the dimensionless compaction viscosity.
The melt mobility number and dimensionless compaction viscosity writes,
\begin{equation}
\label{eq:melt_mob_number}
    \begin{split}
      \delta=\sqrt{\frac{a^2}{C_0 D^2} \frac{\eta_s}{\eta_l}}, \\
    \end{split}
\end{equation}
where $a$ is the crystal size, $C_0$ is a constant. 
Note that in Equation~\eqref{eq:retroaction} we made the assumption that the bulk viscosity or compaction viscosity $\zeta \propto \eta_s^{\rm ref} / \phi$ \cite{bercoricard2001,wallner2016} that expresses the ease at which melt pores in the solid matrix can be deformed or closed. In our model, the reference bulk viscosity  ($\eta_s^{\rm ref}$)  and the shear viscosity ($\eta_s$) of the solid are equal.

On a discretized domain, solving the momentum conservation equation for the two mechanical phases involves the inversion of two sparse matrices, which are the most time-consuming operations. In {\tt Bambari}, these  matrix inversions are performed using the direct  {\tt PARDISO} library that is parallelized using {\tt OpenMP} directives \cite{pardiso-7.2a,pardiso-7.2b}. We use a Cartesian grid resolution of $200\times600$ square cells (aspect ratio is 3:1). The maximum thermal Rayleigh - that would characterize the convective vigor if the numerical domain was fully molten - is $10^9$. Both surface and bottom temperature are imposed and constant. The surface temperature follows the gradient temperature imposed by stellar radiation \cite{nguyen2020}. Mechanical boundary conditions are free-slip on the four sides of the domain.

For phase change, we assume thermodynamic equilibrium between solid and liquid at every time-step, everywhere in the numerical domain. Phase change is thus not limited by reaction kinetics but only by phase diagram (that describes the solid-liquid equilibrium) and energy conservation (see Supplementary Figure 2) \cite{wallner2016}.  We parameterize the mantle melting relations using an idealized binary loop that is fitted on high-pressure diamond anvil-cell experiments (see SI section 1).

In Supplementary Table 1, we report typical estimates of magma oceans physical parameters and associated dimensionless parameters used in the 2D Cartesian simulations.

\subsection{3D turbulent flow model}

In the fully-molten, low viscosity case, we numerically model the magma ocean circulation by solving for turbulent thermal convection in a rotating spherical shell using the open-source code {\tt MagIC} \cite{wicht_inner-core_2002,christensen_numerical_2001,schaeffer_efficient_2013}. We solve the Navier-Stokes, mass conservation and energy conservation equations in spherical coordinates under the Boussinesq approximation \citep{gastine2016}. Consistently with the 2D multiphase flow model, we use free-slip velocity boundary conditions at the bottom and at the top of the ocean. At the bottom surface the temperature is fixed to  $T_{\rm bot}$.  We impose a heterogeneous temperature field at the top of the ocean to represent the day-side and the night-side of the planet:
	\begin{equation}
		T_{\rm top}(\theta,\phi) = T_{0} (1+ \gamma Y_1^1(\theta,\phi)),
	\end{equation}

\noindent where $Y^1_{1}(\theta,\phi)=-1/2 \sqrt{3/2\pi} \sin(\theta) \cos(\phi)$ is the hemispheric spherical harmonic, $\theta$ is the colatitude ($\theta \in [0,\pi]$) $\phi$ is the longitude ($\phi \in[0,2\pi]$) and $T_0$ is the average top temperature. To be consistent with the 2D model, $\gamma$ is chosen such that the hottest temperature, at the sub-stellar point, is equal to the temperature at the bottom of the ocean. The maximum temperature gradient at the surface is expected to be located at 120$^{\circ}$ from the substellar point due to the super-illumination \cite{leger2011,nguyen2020}. Here, we use a simplified configuration (hemispheric spherical harmonic) where the largest temperature gradient is located at 90$^{\circ}$. In {\tt MagIC}, with Dirichlet boundary conditions, the average temperature jump $\Delta T=T_{\rm bot} - T_0$ is used as the temperature scale, and the top and bottom dimensionless temperatures are set to $-r_i^2/(r_i^2+r_o^2)$ and $r_o^2/(r_i^2+r_o^2)$ respectively. In our case, $r_i/r_o = 0.5$ and $r_o-r_i=D=1$, leading to $T_\textup{{bot}}=0.8$, $T_0=-0.2$ and the temperature at the outer boundary varies between $T_{\rm top}^{\rm max}=0.8$ at the sub-stellar point and $T_{\rm top}^{\rm min}=-1.2$ at the anti-stellar point. 

The following governing dimensionless conservation equations are solved in a spherical shell geometry:
	\begin{align}
		\frac{\partial \bm{u}}{\partial t} + \bm{u \cdot \nabla} \bm{u}  &= - \bm{\nabla} p - \frac{2}{E} \bm{e}_z \times \bm{u} + \frac{Ra}{Pr} \frac{r}{r_o} T \bm{e}_r +  \nabla^2 \bm{u}, \label{eq:NS} \\
		\frac{\partial T}{\partial t} +  \bm{u \cdot \nabla} T &= \frac{1}{Pr} \nabla^2 T, \label{eq:energy}\\
		\bm{\nabla \cdot u} &= 0, \label{eq:mass}
	\end{align}
where equation \eqref{eq:NS} accounts for conservation of momentum (Navier-Stokes equation), equation \eqref{eq:energy} accounts for energy conservation and equation \eqref{eq:mass} accounts for mass conservation. Here, $\bm{e}_r$ and $\bm{e}_z$ are the radial and vertical unit vectors, respectively, $T$,  $p$ and $\bm{u}$ are the dimensionless fluid temperature, pressure and velocity, and the gravity field is taken to be linear $\bm{g}=g \bm{r}/r_0$. The solution to the system of equations \eqref{eq:NS}-\eqref{eq:mass} depends on three dimensionless parameters: the Rayleigh number $Ra$, the Ekman number $E$, and the Prandtl number $Pr$:
	\begin{equation}
		\begin{aligned}
			Ra &=&  \frac{\alpha g \Delta T D^3 }{\nu \kappa}, \\
			E  &=&  \frac{\nu}{\Omega D^2}, \\
			Pr &=&  \frac{\nu}{\kappa},
		\end{aligned}
	\end{equation}

\noindent where $\nu=\eta_l/\rho$, is the kinematic viscosity, assumed to be homogeneous in the entire domain.

The system is therefore governed by three dimensionless parameters, the thermal Rayleigh number $Ra$ (buoyancy/diffusion), the Ekman number $E$ (viscosity/rotation) and the Prandtl number $Pr$ (viscosity/thermal diffusivity). In addition, the bulk convective Rossby number $Ro_c$ is a useful proxy to qualitatively assess the rotation influence on the convective flow \citep{gilman1977nonlinear}. In a slowly-rotating regime, it compares the free-fall time of a fluid parcel with a rotation period \citep{aurnou_connections_2020}:
	\begin{equation*}
		Ro_c = \frac{U_\textup{ff}}{\Omega D} = \frac{\sqrt{\alpha g \Delta T D}}{\Omega D} = \frac{Ra^{1/2} E}{Pr^{1/2}},
	\end{equation*}
where $U_\textup{ff}$ is the convective free-fall velocity, $g$ is the gravitational acceleration, $\alpha$ is the thermal expansivity, $D$ is the thickness of the magma ocean, rotating at a rate $\Omega$ about the vertical axis $z$. In Supplementary Table 2, we report typical estimates of magma oceans physical parameters and associated dimensionless parameters. In a low-viscosity scenario, we obtain $Ro_c$ in a range $1.4-7.6$ (Supplementary Table 2). Since the Rossby number could reach values of order one, rotational effects, through the Coriolis force, can be as important as buoyancy effects. Therefore, we argue that both non-rotating and rotating regimes should be considered. Since realistic dimensionless parameters are out-of-reach with current computational capabilities, we work at more moderate Ekman and Rayleigh numbers, chosen such that the convective Rossby number remains of order one ($Pr=1$, $Ra=\num{1.1e8}$, $E=\num{1.5e-4}$ and $Ro_c=1.6$). This ensures the right force balance between buoyancy and rotation. For the non-rotating simulations, the Prandtl and Rayleigh are left unchanged and the Ekman number becomes irrelevant (\textit{i.e.}, $E\longrightarrow\infty$, $Ro_c \longrightarrow \infty$).

{\tt MagIC} employs a pseudo-spectral method with Chebyshev polynomials in the radial direction and spherical harmonics in the longitudinal and latitudinal directions, and the fast spherical harmonic transform library is used {\tt SHTns} \cite{schaeffer_efficient_2013}. The simulations were run with 201 grid points in radius, 1792 grid points in longitude and 896 in latitude. The equations are advanced in time using a mixed time stepping scheme (Crank-Nicolson for the implicit terms and a second-order Adams-Bashforth for the explicit terms). The simulations were run for 350 and 600 turnover times for the non-rotating and rotating simulations respectively, ensuring temporal convergence.  For further details on the numerical methods, we refer the reader to code  documentation (available at \hyperlink{https://magic-sph.github.io/}{https://magic-sph.github.io/}). {\tt MagIC}  is parallelized using both {\tt OpenMP} ( http://openmp.org/wp/) and {\tt MPI} (http://www.open-mpi.org/) and the simulations were performed on the \textit{Lonestar6} nodes of TACC supercomputer (http://www.tacc.utexas.edu).

\subsection{Atmosphere composition and spectroscopy}

The method for generating the emission spectra from Fig~\ref{fig:spectra} has been developed in the framework of another study \citep{falco2024outgassing}.
The chemistry of the atmosphere is assumed to be in equilibrium and is computed by ATMO \citep{Amundsen2014,tremblin2016cloudless}. It does not consider photochemical processes, which are not expected to be important at these high temperatures \cite{2014RSPTA.37230073M}. 
ATMO has been adapted to account for the vapor elemental composition calculated by a gas-liquid equilibrium model, namely MAGMAVOL, introduced in \cite{charnoz2023effect} and used in the methodology of \cite{falco2024outgassing}.
The atmospheric molecular composition is calculated by identifying the point where the pressure-temperature profile intersects the vapor pressure curve calculated using the method of \cite{charnoz2023effect}, which also determines the elemental composition of the vapor.

\subsection{Thermal Mid-IR phase curve}
We first calculate the expected eclipse depth over the MIRI LRS bandpass (5-10.1 um) for some of the most favourable lava planet targets, HD 20329b, K2-141b, HD 3167b, HD 213885b and TOI-431b. Figure 5a shows the expected dayside temperature and uncertainty where we assume a blackbody emission for the host star and the planet, and no recirculation of heat to the night-side hemisphere. We computed the expected planet's emitted flux, $F_p/F_*$, assuming temperatures of 500, 1000, 1400, and 2000 K respectively to estimate the expected phase curve amplitude for different night-side temperatures. We estimate the uncertainty on the broadband MIRI LRS eclipse depth for each target with Pandexo \cite{pandexo_batalha} with an integration time that ensures the saturation level remains under 80\%, a baseline as long as the eclipse duration and a noise floor of 10 ppm. We elect to consider only wavelength below 10.1 $\mu$m as MIRI LRS is known to exhibit a severe instrumental noise along with a large drop in flux at longer wavelength. We compare these uncertainties to the expected phase curve amplitudes to show that this precision is sufficient to distinguish the phase curve amplitude of cold-interior case from a hot-interior case.

To further demonstrate that we can distinguish different internal thermal state with a MIRI LRS phase curve, we simulate phasecurve observations using \texttt{batman} \cite{batman_kreidberg} to model the transit and eclipses and modelled the phase curve modulation as a first order sinusoid with no phase offset. Since the precision on the phase curve amplitude depends on the precision of the measured eclipse depth, we used the eclipse depth uncertainty to represent the photometric uncertainty achieved for temporal bins of the duration of eclipse, i.e. 56.4 minutes.
Figure \ref{fig:phasecurvemiri} shows the simulated phase-folded phase curve of K2-141b for 2 full-orbits and the expected $F_p/F_*$ assuming a cold interior with a $T_{night}=$ 500 K. We then fit the simulated data using \texttt{emcee} \cite{2013PASP..125..306F}, a Markov-Chain Monte Carlo (MCMC) routine, with 100 walkers and 5000 steps each, where we let the eclipse depth and phase curve amplitude to vary freely and show that our simulated MIRI LRS phase curve would is sufficiently precise to distinguish a cold interior from a hot interior phase curve (see section 9 in supplemental information for more details).


\section{Data availability}

All data used in this manuscript are accessible via the fair depository of the Institut de Physique du Globe de Paris \href{https://doi.org/10.18715/IPGP.2024.m41y3glp}{https://doi.org/10.18715/IPGP.2024.m41y3glp}.

\section{Code availability}

The code {\tt MagIc} \cite{wicht_inner-core_2002,christensen_numerical_2001,schaeffer_efficient_2013} is open-source and can be accessed at \hyperlink{https://magic-sph.github.io/}{https://magic-sph.github.io/}). {\tt MagIC} version 6.0 was used and is also available on Zenodo: https://zenodo.org/records/4590225. The code {\tt Bambari} can be accessed on GitHub upon reasonable request. 

\section{Acknowledgments}

This work has received funding from the European Research Council (ERC) under the European Union's Horizon 2020 research and innovation program (grant agreement no. 101019965— SEPtiM). Parts of this work were supported by the UnivEarthS Labex program at Université de Paris and IPGP (ANR-10-LABX-0023 and ANR-11-IDEX-0005-02) and Natural Sciences and Engineering Research Council of Canada (RGPIN-2024-06174).
Two-dimensional numerical computations were performed on the IPGP S-CAPAD/DANTE platform. D. Lemasquerier acknowledges the Texas Advanced Computing Center (TACC) at The University of Texas at Austin for providing HPC and visualization resources that have contributed to the research results reported within this paper (URL: http:// www.tacc.utexas.edu). L.D.\ acknowledges support from the Banting Postdoctoral Fellowship program, administered by the Government of Canada and the Trottier Family Foundation.

\section{Author contributions}

\textbf{Author 1: C.-E. Boukaré } - Conceived and designed the analysis - Designed the numerical simulations -  Performed the numerical simulations - Produced the figures - Wrote the manuscript. \textbf{Author 2: D. Lemasquerier} Performed the numerical simulations - Produced the figures and analyzed the data - Revised the manuscript. \textbf{Author 3: N. Cowan} Conceived and designed the analysis - Wrote the manuscript. \textbf{Author 4: H. Samuel} Designed the numerical simulations - Revised the manuscript. \textbf{Author 5: J. Badro} Conceived the analysis - Revised the manuscript.  \textbf{Author 6: L. Dang} Performed the analysis - Produced the figures. \textbf{Author 7: A. Falco} Performed the analysis - Produced the figures. \textbf{Author 6: S. Charnoz} Performed the analysis - Revised the manuscript.   

\section{Competing interests}

The authors declare no competing interests. 

\section{Additional information}

Physical parameters used in the models can be found in the Supplementary Information. The videos corresponding to Figure~3 are also included in the supplementary information.

\newpage

\newpage

\FloatBarrier



\end{document}